\documentclass[twocolumn]{aastex701}

\usepackage{bm}
\usepackage[export]{adjustbox}
\usepackage{mathrsfs,amsmath,amssymb}

\def\apjl{Astrophys. J. Lett.}
\def\apjs{Astrophys. J., Suppl. Ser.}

\def\aap{Astron. Astrophys.}

\def\jcap{J. Cosmology Astropart. Phys.}

\begin{document}

\title{Listening to Black Mirrors With Gravitational Radiation}

\author[orcid=0000-0003-3993-3249,gname='Pau',sname='Amaro Seoane']{Pau Amaro Seoane}
\affiliation{Universitat Politècnica de València, Spain}
\affiliation{Max Planck Institute for Extraterrestrial Physics, Garching, Germany}

\email[show]{amaro@upv.es}

\begin{abstract}
The existence of curvature singularities and the information and firewall paradoxes are significant problems for the conventional black hole model. The black mirror provides a CPT-symmetric alternative to the classical description. We show that classical black holes can be distinguished from black mirrors by using gravitational waves. The principal challenge is to identify a unique, testable signature of the black mirror's reflective horizon that can be detected. The horizon singularity of the black mirror model necessitates that no energy flux is propagated beyond the horizon, which can be described effectively by imposing specific boundary conditions at the event horizon. We demonstrate that the quasi-normal mode spectrum of the black mirror is fundamentally different from that of classical black holes. We derive the reflectivity of the black mirror and find it is given precisely by the generalized Boltzmann factor. 
Moreover, we show that this is a universal behaviour: regardless of the specific details of the unknown quantum gravity interactions, the macroscopic reflectivity is dictated solely by the Hawking temperature $T_H$. This drastically alters the orbital dynamics of extreme-mass ratio inspirals. For low spins, the inspiral decelerates due to reduced absorption. For high spins and prograde orbits, the black mirror suppresses the superradiant amplification of classical black holes, acting instead as an absorber. This leads to an inspiral that proceeds faster than the classical prediction. Finally, we show that this model allows for the cosmic growth of supermassive black holes to high spins via accretion. A definitive detection of these signatures would provide compelling evidence distinguishing the reflective boundary of a black mirror from the perfectly absorbing horizon of a classical black hole.
\end{abstract}

\section{Introduction}

\begin{figure}
  {\includegraphics[width=0.45\textwidth,center]{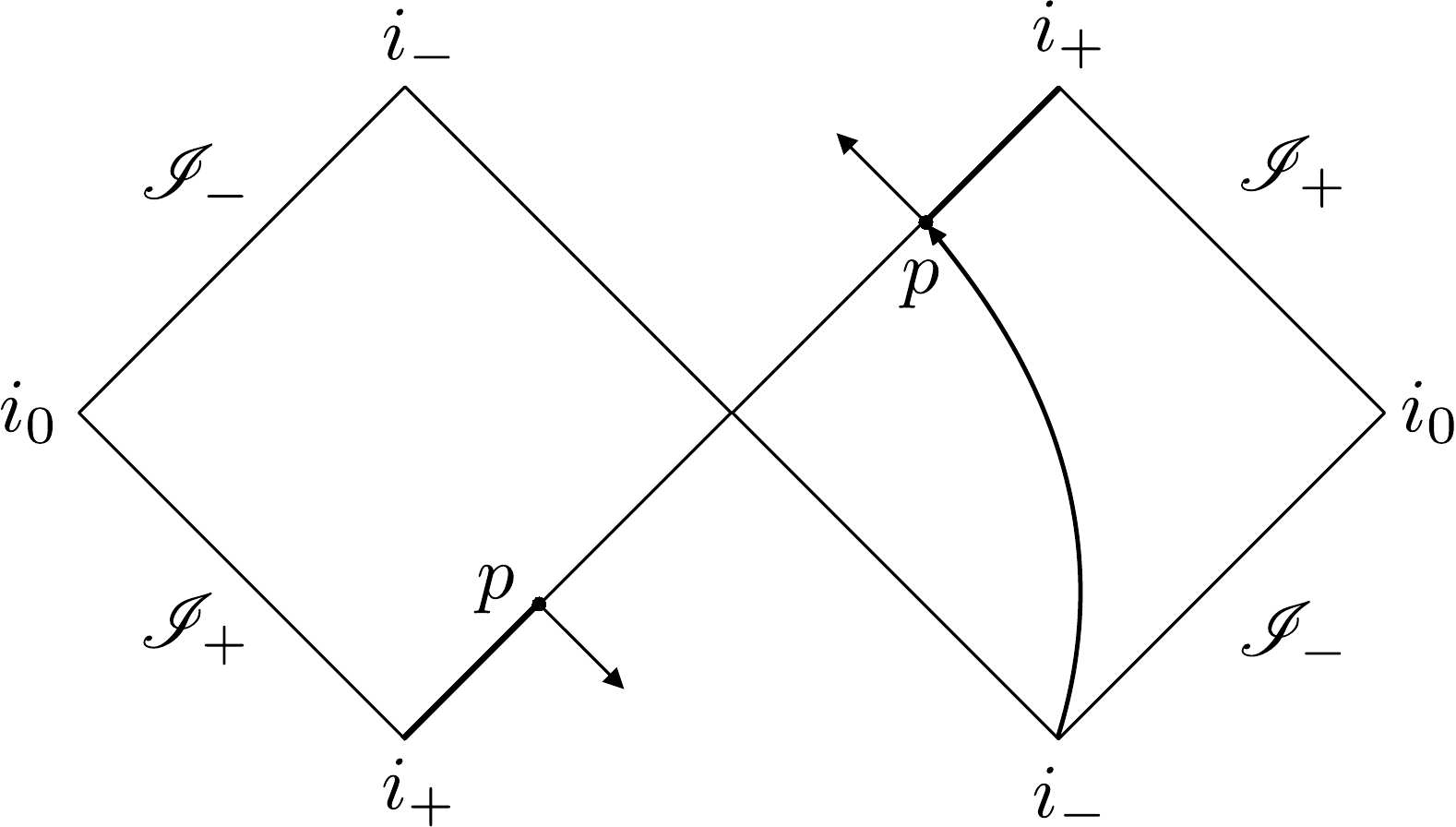}}
\caption
{
A massive particle intersecting the horizon of the black mirror at a point $p$ is subsequently trapped in the horizon.
An interaction at the horizon results in an outgoing signal, the thicker black segment, which propagates along the horizon. The points $i^+$, $i_-$, and $i_0$ represent future timelike, past timelike, and spacelike infinity, respectively, while $\mathscr{I}_-$ represents past null infinity.
}
\label{fig:black-mirror-penrose}
\end{figure}

The concept of the black mirror \cite{TzanavarisEtAl2024} has recently been put forward as a theoretical alternative to the classical black hole. It is derived as the saddle point of the total gravitational action, under the imposition of a reflection symmetry that the authors have identified as a classical analogue of $CPT$ for the gravitational field, which identifies antipodal points and transversal directions on their horizons. Consequently, if a particle intersects the event horizon, its velocity is always pointing inwards towards the interior. However, as there is no interior, the particle is subsequently confined to the event horizon, and is therefore accelerated to the speed of light.\footnote{This phenomenon of a massive particle becoming massless at the horizon should be interpreted as a collision of the particle and its $CPT$-mirror at the same point on the horizon, resulting in pair annihilation (c.f. \cite{TzanavarisEtAl2024}).}. We illustrate this in Fig.~(\ref{fig:black-mirror-penrose}), a Penrose diagram depicting the spacetime geometry of a black mirror. A massive particle intersecting the horizon of the black mirror at a point $p$ is subsequently trapped, and its quantum information is absorbed and non-locally scrambled by the mirror's internal degrees of freedom. The diagram thus provides a schematic for how information is processed by the mirror rather than being lost.

It is important to note that this picture is inherently classical, and most likely breaks down near the horizon. Nevertheless, it does provide a recipe for an effective description of the boundary conditions one must impose to describe the propagation of gravitational waves in this background. The fact that the horizon acts much like a wall, preventing anything from falling into an interior (as there is no interior for something to fall into) means that the energy flux of any wave cannot penetrate the horizon. This condition, or variants of it, appear in similar models such as the membrane paradigm \cite{thorne1986membrane,jacobson2017membrane} and black hole complementarity \cite{Susskind:1993mu,Almheiri:2012rt}. Therefore, if $\psi$ is a scalar linear wave, then $\psi$ must be constant in the vicinity of the event horizon. The imposition of this boundary condition has profound implications.

Specifically, we show that the black mirror hypothesis predicts key deviations from classical general relativity: (1) that the quasi-normal mode spectrum is altered; (2) that the reflectivity of the near-horizon structure is given by the generalized Boltzmann factor {which depends on the absolute value of the co-rotating frequency}; and (3) that this reflectivity leads to observable consequences in gravitational wave signals, such as modified inspiral dynamics. Our analysis shows the black mirror's reflectivity is a universal result, independent of the specific quantum dynamics at the horizon.

Long inspirals of asymmetric binaries, such as extreme- or intermediate-mass ratio inspirals (EMRIs) \citep{AmaroLRR,Amaro2007,AmaroHandbook} and extremely-large mass ratio inspirals (XMRIs) \citep{amaro2019extremely}, are ideal for testing this idea, as their sustained gravitational wave signals act as a continuous probe of the central object's near-horizon structure. An EMRI, IMRI, or XMRI provides a continuous probe of the near-horizon geometry for up to hundreds of thousands of orbits. This secular process allows for the accumulation of phase shifts, tests that are impractical with shorter signals. This method is particularly powerful because it can reveal how the horizon's reflectivity changes with the wave's frequency and the central object's spin, providing a rich data set to test the underlying physics of the black mirror.

A space-borne gravitational wave observatory, such as LISA, TianQin, or Taiji \cite{LISA2017,TianQin2025,Taiji2020}, would be ideally positioned to measure the subtle signatures of black mirrors. These observatories are designed to detect gravitational waves in the millihertz frequency band, which is the precise range where EMRIs and XMRIs are expected to operate. Their long-term observation capabilities and unprecedented sensitivity would allow them to distinguish between a classical black hole and a black mirror.

The detection or non-detection of these signatures would not merely be an incremental finding but a crucial discriminator between competing theories of gravitational collapse and the fundamental nature of spacetime itself.

\section{The quasi-normal modes}

Linear perturbations of the exterior of a non-rotating black hole of mass $M$ are described by a wave equation of the form 
\begin{equation}
(-\partial_t^2 + \partial_x^2 - V_\ell(r))\psi_\ell = 0,
\end{equation}
where $x = r + 2M\ln(r/2M-1)$ is the \textit{tortoise coordinate}, and $V_\ell$ is the Regge-Wheeler/Zerilli potential \cite{regge1957stability,zerilli1970effective}. We adopt the standard convention in General Relativity and seek monochromatic solutions of the form $\psi_{\ell}=e^{-i\omega t}\psi_{\ell,\omega}$. The \textit{quasi-normal modes} (QNMs) are solutions of specific (complex) frequency $\omega$.

For a classical black hole, QNMs must satisfy specific boundary conditions: purely ingoing waves at the horizon ($x\to -\infty$) and purely outgoing waves at spatial infinity ($x\to+\infty$). Under the $e^{-i\omega t}$ convention, an ingoing wave has the spatial form $e^{-i\omega x}$, and an outgoing wave has the spatial form $e^{+i\omega x}$. Consequently, the spatial part $\psi_{\ell,\omega}$ must satisfy the boundary value problem
\begin{subequations}
\begin{align}
& (\partial_x^2 + \omega^2 - V_\ell(r))\psi_{\ell,\omega} = 0,
\label{eqn:quasi-equation}\\
& \psi_{\omega,\ell}(x)\to e^{+i\omega x} \quad\textrm{as}\quad x\to+\infty, \\
& \psi_{\omega,\ell}(x)\to e^{-i\omega x} \quad\textrm{as}\quad x\to-\infty.
\end{align}
\end{subequations}
As discussed in the previous section, in the black mirror case, the boundary conditions in the near-horizon limit change, as one must impose the no-flux condition, which in our case is
\begin{equation}
\label{eqn:no-flux-cond}
\psi_{\ell,\omega}\to\textrm{constant}\quad \textrm{as} \quad x\to -\infty.
\end{equation}
Following \cite{chandra1975quasi}, suppose that there is a function $\phi_{\ell,\omega}$ such that $\psi_{\ell,\omega}(x) = \exp\left(i\int_0^x\phi_{\ell,\omega}(x')\,dx'\right)$. Substituting this ansatz into eq. \eqref{eqn:quasi-equation}, it follows that $\phi_{\ell,\omega}$ must satisfy a Ricatti-type equation of the form 
\begin{equation}
\label{eqn:quasi-ricatti}
\partial_x\phi_{\ell,\omega} + i(\phi_{\ell,\omega}^2 + V_\ell(r) - \omega^2) = 0.
\end{equation}
To find the modes of the black mirror, we impose the boundary conditions corresponding to an outgoing wave at infinity and the no-flux condition at the horizon,
\begin{equation}
\lim_{x\to+\infty}\phi_{\ell,\omega}(x)=+\omega,
\quad
\lim_{x\to-\infty}\phi_{\ell,\omega}(x)= 0.
\end{equation}
The last boundary condition guarantees that $\psi_{\ell,\omega}$ is constant in the near-horizon limit. However, applying this boundary condition to eq. \eqref{eqn:quasi-ricatti} (noting that $V_\ell \to 0$ as $x\to-\infty$) yields $i(0 + 0 - \omega^2) = 0$, which implies $\omega = 0$. Thus, the resulting solution represents a stationary perturbation and not a propagating gravitational wave. In other words, \textit{the quasi-normal mode spectrum of a black mirror is fundamentally different, as modes satisfying the standard QNM boundary conditions combined with the black mirror condition do not exist.}

\section{The reflectivity of black mirrors}

The alteration of the QNM spectrum, which follows as a direct consequence of the effective description of the no-flux condition \eqref{eqn:no-flux-cond}, means that the propagation of waves in a black mirror is substantially different from that in standard black holes. Nonetheless, the treatment has been entirely classical up to the horizon, where we know that quantum effects must dominate in the near-horizon limit. To determine the interaction of gravitational waves with the black mirror, we must incorporate these quantum effects.

In the near-horizon limit, the solution to the wave equation \eqref{eqn:quasi-equation} is a superposition of ingoing ($e^{-i\omega x}$) and outgoing ($e^{+i\omega x}$) components,
\begin{equation}
    \psi_{\omega,\ell}(x) \simeq A_{\text{in}} e^{-i\omega x} + A_{\text{out}} e^{i\omega x}.
\end{equation}
\noindent
The energy reflectivity $R$ is defined as the ratio of the outgoing energy flux to the ingoing energy flux, $R(\omega) = |A_{\text{out}}|^2 / |A_{\text{in}}|^2$.

{Physically, the horizon interaction arises from the coupling to a dissipative channel. A purely viscous term in the effective wave equation generates dissipation but cannot source energy. As shown by \cite{OshitaEtAl2020} (see Eqs.~2.13--2.16 therein), the reflectivity of the horizon is not an arbitrary choice but a property of the exact solution in Rindler coordinates. This derivation explicitly demonstrates that the dissipation term depends on the magnitude of the frequency relative to the thermal scale.}

{Consequently, the reflectivity is given by the Boltzmann factor involving the absolute value of the frequency,}
\begin{equation}
\label{eq.R_Schwarzschild_Boltzman}
R(\omega) = e^{-|\omega|/T_H}.
\end{equation}
{We emphasize that the absolute value in the exponent is not posited ad-hoc. It is a necessary consequence of the requirement that the horizon response is causal and stable ($R \le 1$). Without the absolute value, modes with negative frequency (which appear in the co-rotating frame of a spinning black hole) would lead to exponential amplification ($R > 1$), violating the premise of a passive, viscous boundary.}

This result is entirely independent of the specific, unknown details of the microscopic quantum gravity interactions at the horizon. Consequently, the black mirror reflects incident radiation with a perfect thermal spectrum.

\section{Observational Distinguishability and Universality}

The fundamental distinction between a classical black hole and a black mirror
resides in the boundary conditions imposed at the event horizon. In the
classical paradigm, the horizon is a perfectly absorbing one-way membrane.
Mathematically, under the $e^{-i\omega t}$ convention, this is realized by requiring solutions to the perturbation
equation \eqref{eqn:quasi-equation} to be purely ingoing in the near-horizon
limit ($x\to-\infty$), specifically $\psi_{\ell,\omega}(x) \propto \exp(-i\omega
x)$. This condition dictates that all incident energy flux propagates into the
black hole interior. In contrast, the black mirror hypothesis, characterized by
the absence of an interior, imposes the no-flux condition
\eqref{eqn:no-flux-cond}. This fundamentally
alters the quasi-normal mode spectrum and
implies that the horizon must possess reflective properties.

The interaction of gravitational waves with this structure is governed by the requirement that the horizon behaves as a quantum system in thermal equilibrium. This leads directly to the reflectivity being given precisely by the Boltzmann factor, {$R = \exp(-|\omega|/T_H)$}.

This result signifies a universal behaviour: regardless of the specific details of the unknown quantum gravity interactions, the macroscopic reflectivity is dictated solely by the Hawking temperature $T_H$. This leads to an
observational degeneracy, as a distant observer cannot passively distinguish
between a classical black hole that absorbs perfectly and subsequently emits
thermal Hawking radiation, and a black mirror that reflects thermally according
to the Boltzmann factor.

\section{Detectability with extreme-mass ratio inspirals}

The detectability of the black mirror signature via EMRI dephasing depends on the magnitude of the reflectivity in the frequency band probed by the inspiral. This depends strongly on the mass $M$ and the spin $a$ of the central object, and critically on the orbital configuration, including inclination.

\subsection{Schwarzschild Case: Frequency Scales}

We first review the non-rotating (Schwarzschild, $a=0$) case. We work in geometric units ($G=c=1$) and natural units ($\hbar=k_B=1$). We consider a non-rotating black hole of mass $M$. The Hawking temperature $T_H$ is defined in terms of the surface gravity $\kappa$ via the relation $T_H = \kappa/(2\pi)$ \cite{Hawking:1975vcx,Gibbons:1976pt,gibbons:1977action}. The Hawking temperature of a Schwarzschild black hole is a function only of its mass $M$, and is given by 
\begin{equation}
\label{eq.TH_Schwarzschild}
T_H = \frac{1}{8\pi M}.
\end{equation}
\noindent
We analyze a test particle in a circular orbit at a Schwarzschild radial coordinate $r$. The angular velocity $\Omega$ of the orbit is determined by the geodesic equation, yielding the relativistic generalization of Kepler's third law,
\begin{equation}
\label{eq.Omega_Schwarzschild}
\Omega = \sqrt{\frac{M}{r^3}}.
\end{equation}
\noindent
In the quadrupole approximation, the dominant gravitational wave emission occurs at twice the orbital frequency. This physical frequency $\omega$ is positive by definition,
\begin{equation}
\label{eq.omega_GW}
\omega = 2\Omega = 2\sqrt{\frac{M}{r^3}}.
\end{equation}
\noindent
We calculate the normalized frequency $\omega/T_H$ by dividing Eq.~(\ref{eq.omega_GW}) by Eq.~(\ref{eq.TH_Schwarzschild}).
Rearranging the terms yields the expression dependent only on the compactness $M/r$,
\begin{equation}
\label{eq.Normalized_Freq_Radius_Derived}
\frac{\omega}{T_H} = 16\pi \left(\frac{M}{r}\right)^{3/2}.
\end{equation}
\noindent
The inspiral phase terminates at the Innermost Stable Circular Orbit (ISCO). For a Schwarzschild black hole, the ISCO is located at $r_{\text{ISCO}} = 6M$. We evaluate the normalized frequency at the ISCO by substituting this radius into Eq.~(\ref{eq.Normalized_Freq_Radius_Derived}),
\begin{equation}
\frac{\omega_{\text{ISCO}}}{T_H} = \frac{8\pi}{3\sqrt{6}}.
\end{equation}
\noindent
The energy reflectivity $R$ at this frequency is given by the Boltzmann factor (Eq.~\ref{eq.R_Schwarzschild_Boltzman}). Since the observed astrophysical frequency is positive ($\omega>0$),
\begin{equation}
R(\omega_{\text{ISCO}}) = e^{-|\omega_{\text{ISCO}}|/T_H} = e^{-{8\pi}/({3\sqrt{6})}} \approx 0.0327.
\end{equation}
\noindent
The EMRI probes the transition from significant reflectivity at larger radii to low reflectivity near the ISCO. The reduced absorption $(1-R(\omega))$ slows down the inspiral compared to a classical black hole.

\subsection{Kerr Case: Spin, Thermodynamics, and Superradiance}

The generalization to the Kerr geometry requires incorporating rotational effects into the thermodynamic description of the horizon. A Kerr black hole is characterized by its Hawking temperature $T_H(M, a)$ and the angular velocity of its horizon, $\Omega_H(M, a)$ \citep{Bardeen1973CMaPh..31..161B}. The interaction between gravitational waves and the rotating horizon depends on the specific wave mode $(\omega, m)$. The relevant frequency governing the interaction is the frequency measured in the co-rotating frame of the horizon \citep{Starobinsky1973JETP...37...28S, Teukolsky1973ApJ...185..635T},
\begin{equation}
\label{eq.corotating_frequency}
\tilde{\omega} = \omega - m\Omega_H.
\end{equation}
\noindent
The reflectivity of the black mirror is determined by the requirement that the horizon behaves as a quantum system in thermal equilibrium. Physical dissipation arises from a viscous term which acts to extract energy from the perturbation. Therefore, the thermodynamically consistent reflectivity that respects causality and stability for a passive boundary is given by
\begin{equation}
\label{eq.Kerr_Reflectivity}
R(\omega, m) = e^{-|\tilde{\omega}|/T_H}.
\end{equation}
\noindent
{This form has significant implications for the physics of superradiance. In classical general relativity, the event horizon amplifies modes with $\tilde{\omega} < 0$ (superradiance), effectively behaving as if it had a reflectivity $R_{\text{class}} > 1$. This amplification extracts rotational energy from the black hole and pumps it back into the orbital environment.}

{In contrast, the black mirror hypothesis, constrained by the requirement $R \le 1$, suppresses this amplification. For superradiant modes, where $|\tilde{\omega}|$ is non-zero, the reflectivity $R = e^{-|\tilde{\omega}|/T_H}$ is strictly less than 1. In fact, for high spins where $T_H$ is small, $R \approx 0$. The black mirror acts as a perfect absorber in the regime where a classical black hole acts as an amplifier.}

The modification to the energy balance during an EMRI is thus fundamentally different from classical expectations. The absorption probability is $\Gamma(\omega, m) = 1 - R(\omega, m)$. {For superradiant modes, $\Gamma \approx 1$ (absorption). This represents a net loss of energy from the orbit into the horizon, whereas in the classical case, the energy flow would be from the horizon to the orbit ($\Gamma_{\text{class}} < 0$).}

The net effect on the orbital evolution depends on the balance between the energy radiated to infinity $\dot{E}_{\infty}$ and the energy absorbed by the horizon. The rate of change of the orbital energy $E_{\text{orb}}$ is $\dot{E}_{\text{orb}} = -\dot{E}_{\infty} - \Gamma \dot{E}_{\text{inc}}$, where $\dot{E}_{\text{inc}}$ is the incident flux near the horizon.

{In the superradiant regime, the black mirror absorbs the incident flux ($\Gamma > 0$), increasing the total energy loss rate of the binary. Conversely, a classical black hole would return energy ($\Gamma < 0$), decreasing the total loss rate. Therefore, we predict that EMRIs around rapidly spinning black mirrors will inspiral \textit{faster} than their classical counterparts.}

\subsubsection{Recovering the Schwarzschild limit}

\noindent
We verify the self-consistency of the generalized Boltzmann factor derived for the Kerr geometry (Eq.~\ref{eq.Kerr_Reflectivity}) by examining its behavior in the limit of zero rotation ($a_* \to 0$). A consistent model must reproduce the results obtained for the Schwarzschild case (Eq.~\ref{eq.R_Schwarzschild_Boltzman}).

We analyze the limits of the horizon angular velocity $\Omega_H$ and the Hawking temperature $T_H$ as $a_* \to 0$.
The horizon angular velocity is defined as \citep{Bardeen1972ApJ...178..347B}
\begin{equation}
\label{eq.derivation_Omega_H}
\Omega_H(a_*) = \frac{1}{M} \frac{a_*}{2(1+\sqrt{1-a_*^2})}.
\end{equation}
Taking the limit $a_* \to 0$, we obtain $\lim_{a_* \to 0} \Omega_H(a_*) = 0$.

\noindent
The Hawking temperature is defined as \cite{Bardeen1973CMaPh..31..161B}
\begin{equation}
\label{eq.consistency_T_H}
T_H(a_*) = \frac{1}{4\pi M} \frac{\sqrt{1-a_*^2}}{(1+\sqrt{1-a_*^2})}.
\end{equation}
\noindent
Taking the limit $a_* \to 0$, we obtain $\lim_{a_* \to 0} T_H(a_*) = \frac{1}{8\pi M} = T_{H, \text{Schw}}$.

We now examine the limit of the reflectivity $R(\omega, m)$. As $a_* \to 0$, the co-rotating frequency approaches the coordinate frequency, $\tilde{\omega} \to \omega$. Since the wave frequency $\omega$ is positive, $|\tilde{\omega}| \to |\omega| = \omega$.

\noindent
Substituting these limits into the generalized Boltzmann factor, we find
\begin{align}
\lim_{a_* \to 0} R(\omega, m) = e^{-\omega/T_{H, \text{Schw}}}.
\end{align}
\noindent
This expression is identical to the reflectivity formula derived for the Schwarzschild case (Eq.~\ref{eq.R_Schwarzschild_Boltzman}). The generalized form is therefore consistent.

\subsection{A Fundamental Threshold}

We analyze the condition for the onset of superradiance for a rotating Kerr black hole with mass $M$ and dimensionless spin parameter $a_*=a/M$. Superradiance occurs when the co-rotating frequency $\tilde{\omega} = \omega - m\Omega_H$ is negative, requiring $\omega < m\Omega_H$ \citep{Brito2015CQGra..32f4001B}. We seek the critical spin $a_c$ at which the dominant mode ($m=2$) for a prograde, equatorial, quasi-circular orbit becomes superradiant exactly at the ISCO.

The gravitational wave frequency is $\omega_{\text{ISCO}} = 2\Omega_{\text{ISCO}}$. The critical condition $\tilde{\omega}=0$ implies
\begin{equation}
\label{eq.derivation_critical_condition}
\Omega_{\text{ISCO}}(a_c) = \Omega_H(a_c).
\end{equation}
\noindent
We utilize the exact expressions for $\Omega_H$ (Eq.~\ref{eq.derivation_Omega_H}) and the orbital frequency $\Omega$ in the Kerr metric \citep{Bardeen1972ApJ...178..347B}. The angular velocity of a prograde equatorial circular orbit at radius $r$ is
\begin{equation}
\label{eq.derivation_Omega_Kerr}
\Omega(r, a_*) = \frac{\sqrt{M}}{r^{3/2} + a\sqrt{M}} = \frac{1}{M} \frac{1}{(r/M)^{3/2} + a_*}.
\end{equation}
\noindent
The orbital frequency at the ISCO, $\Omega_{\text{ISCO}}(a_*)$, is obtained by evaluating Eq.~(\ref{eq.derivation_Omega_Kerr}) at the radius of the prograde ISCO, $r_{\text{ISCO}}(a_*)$. The ISCO radius is given by \citep{Bardeen1972ApJ...178..347B}
\begin{equation}
\label{eq.derivation_rISCO}
r_{\text{ISCO}}(a_*)/M = 3 + Z_2 - \sqrt{(3-Z_1)(3+Z_1+2Z_2)},
\end{equation}
\noindent
where the auxiliary variables $Z_1$ and $Z_2$ are defined as
\begin{align}
Z_1 &= 1 + (1-a_*^2)^{1/3} \left[ (1+a_*)^{1/3} + (1-a_*)^{1/3} \right], \\
Z_2 &= \sqrt{3a_*^2 + Z_1^2}.
\end{align}
\noindent
The critical spin $a_c$ is the solution to the transcendental equation formed by substituting these expressions into Eq.~(\ref{eq.derivation_critical_condition}). We rewrite the condition as
\begin{equation}
\frac{1}{(r_{\text{ISCO}}(a_c)/M)^{3/2} + a_c} = \frac{a_c}{2(1+\sqrt{1-a_c^2})}.
\end{equation}
The resulting equation is transcendental and must be solved numerically. The solution yields the critical spin value,
\begin{equation}
a_c \approx 0.359403.
\end{equation}
\noindent
For spins $a_* > a_c$, the prograde ISCO frequency is lower than the horizon frequency ($\Omega_{\text{ISCO}} < \Omega_H$), satisfying the superradiance condition for the dominant $m=2$ mode.

{Under the stabilized black mirror hypothesis ($R = e^{-|\tilde{\omega}|/T_H}$), this critical spin $a_c$ represents the point of maximum reflectivity. At $a_c$, $\tilde{\omega} = 0$, so $R = e^0 = 1$. The horizon acts as a perfect mirror.}

{For $a_* < a_c$ (low spin), $|\tilde{\omega}| > 0$, so $R < 1$ (partial absorption). For $a_* > a_c$ (high spin), $|\tilde{\omega}| > 0$, so $R < 1$ (partial absorption). The transition across $a_c$ is smooth in terms of stability, but distinct in terms of the energy flux budget.}

\subsection{Two Limiting Regimes}

To illustrate these effects, we present simplified simulations comparing the evolution of EMRIs around classical black holes (BH) and black mirrors (BM). In the quadrupole approximation for circular orbits, the physical strain polarization $h_+$ is given by $h_+ \approx (2\mu/D_L) x^2 \cos(\Phi(t))$, where $x:=M/r$ is the frequency parameter and $\Phi(t)$ is the accumulated phase. We define the normalized strain as $h_{\text{norm}} = x^2 \cos(\Phi)$. 

\noindent
\textit{Low-Spin Regime ($a<a_c$).}---We first examine the low-spin regime. Here $\tilde{\omega} > 0$, so $|\tilde{\omega}| = \tilde{\omega}$. The reflectivity is $R = e^{-\tilde{\omega}/T_H} < 1$.
The absorption probability is $\Gamma = 1-R$. Since $R > 0$, the absorption $\Gamma < 1$.
Compared to a classical Schwarzschild black hole (perfect absorber, $\Gamma=1$), the black mirror absorbs less energy. This leads to a deceleration of the inspiral.

We simulate a typical LISA EMRI system with $M=10^6 M_\odot$ and $\mu=10 M_\odot$, starting from $r=10M$ until the classical BH reaches the ISCO ($r=6M$).

\begin{figure}
 {\includegraphics[width=0.45\textwidth,center]{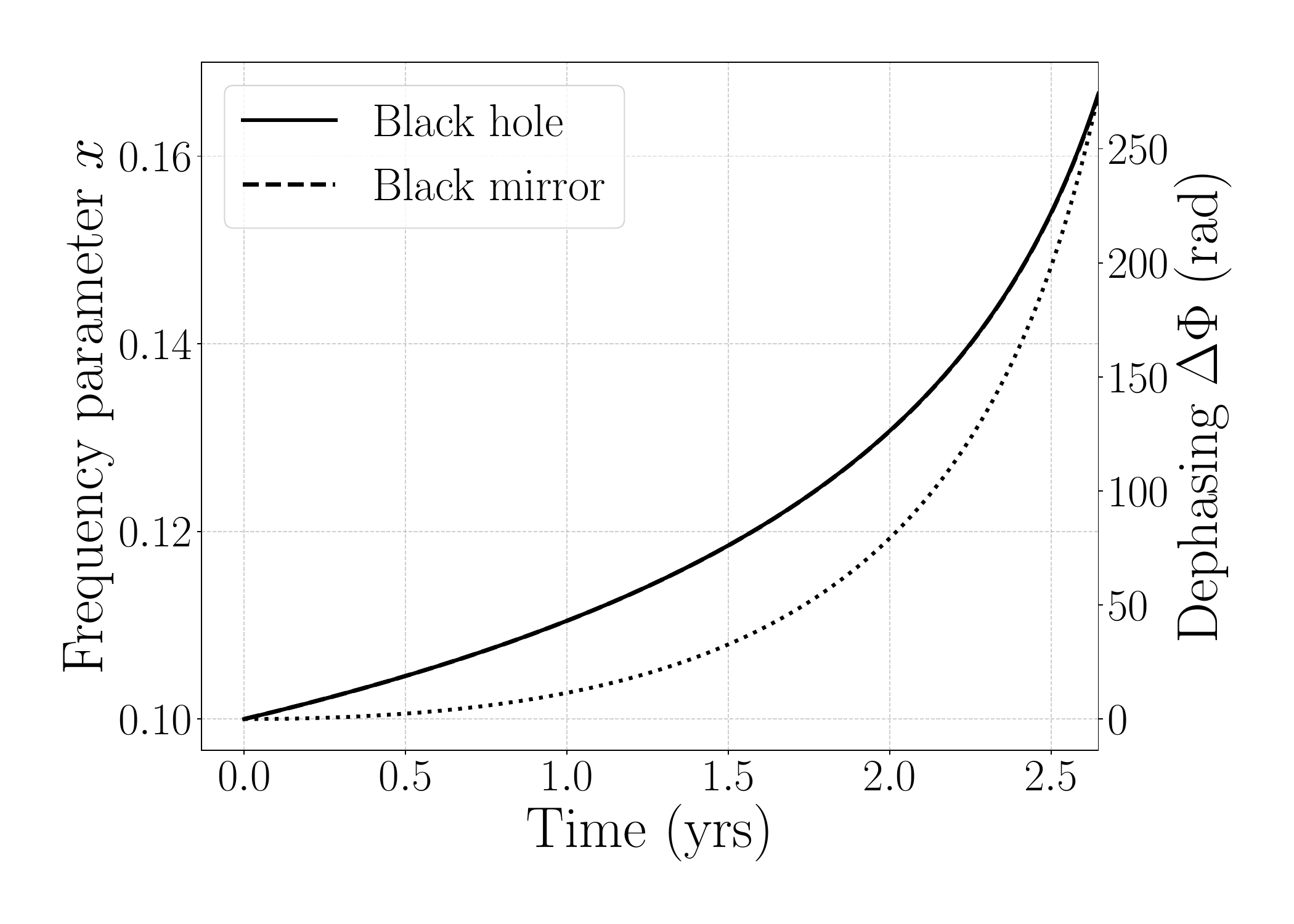}}
\caption
{
EMRI evolution in the low-spin regime ($a_* < a_c$). The evolution of the frequency parameter $x=M/r$ (solid line for BH, dashed line for BM) and the cumulative dephasing $\Delta \Phi$ (dotted line) are shown. The BM inspiral is slower than the BH inspiral due to reduced horizon absorption.
}
\label{fig:BM_Evolution_Dephasing}
\end{figure}

Fig.~(\ref{fig:BM_Evolution_Dephasing}) illustrates the evolution. The figure shows the evolution of the frequency parameter $x$ (left axis) and the accumulated dephasing $\Delta \Phi$ (right axis). The black mirror inspiral (dashed line) is slower than the classical inspiral (solid line). This deceleration occurs because the reflectivity reduces the energy flux absorbed by the horizon. The difference in the evolution of $x$ is small (relative difference $\approx 0.0005$), causing the lines to overlap significantly. However, the effect on the phase is cumulative. The dephasing (dotted line) accumulates secularly over the observation time (approximately 2.65 years), reaching approximately 274 radians.

\begin{figure*}
 {\includegraphics[width=1\textwidth,center]{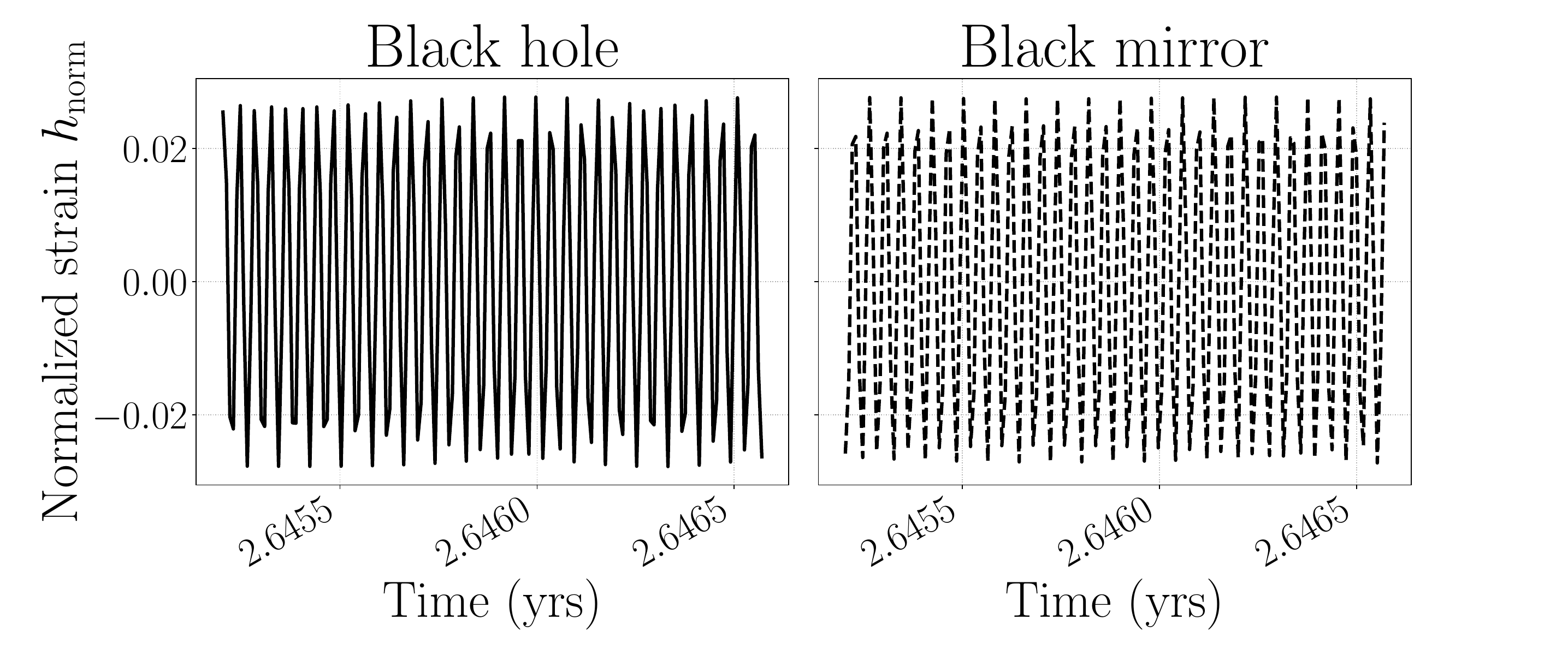}}
\caption
{
Zoom view of the normalized gravitational waveform $h_{\text{norm}}$ in the low-spin regime near ISCO. Left panel: Classical black hole (BH). Right panel: Black mirror (BM). The BM waveform lags behind the BH waveform due to the accumulated dephasing shown in Fig.~(\ref{fig:BM_Evolution_Dephasing}).
}
\label{fig:BM_Waveform_Zoom}
\end{figure*}

Fig.~(\ref{fig:BM_Waveform_Zoom}) presents a zoom view of the normalized gravitational waveform $h_{\text{norm}}$ during the final 0.5 days of the simulation near the ISCO. The left panel shows the classical BH waveform (solid line), and the right panel shows the BM waveform (dashed line). The cumulative effect of the slower inspiral rate is visible; the BM waveform lags behind the BH waveform.

We quantify the effect for $a_*=0.1$. The relevant parameters are $\tilde{\omega} M \approx 0.096$ and $T_H M \approx 0.040$. The normalized frequency is $|\tilde{\omega}|/T_H \approx 2.45$.
The reflectivity is $R \approx e^{-2.45} \approx 0.087$.
The modification to the total flux leads to a dephasing of $\Delta \Phi \approx 1372$ radians over $10^5$ cycles. This confirms that for low spins, the black mirror is distinguishable via a ``slowing down'' of the inspiral.

\noindent
\textit{High-Spin Regime ($a>a_c$).}---We contrast this with a rapidly rotating system, assuming $a_*=0.99$. The dynamics are dominated by the near-extremal geometry. The Hawking temperature is significantly suppressed, $T_H \approx 0.00984/M$. The ISCO is at $r_{\text{ISCO}} \approx 1.454 M$ \citep{Bardeen1972ApJ...178..347B}.

The dominant mode is superradiant ($\tilde{\omega} < 0$), but due to the absolute value in our formula, the reflectivity remains bounded.
The normalized frequency is $|\tilde{\omega}|/T_H \approx 14.14$.
The reflectivity is $R = e^{-14.14} \approx 7 \times 10^{-7} \approx 0$.

{In this regime, the black mirror acts as a nearly perfect absorber ($\Gamma \approx 1$). This stands in contrast to a classical Kerr black hole. For a classical BH with $a_*=0.99$, the superradiant mode is amplified. The horizon effective reflectivity is $R_{\text{class}} > 1$. This implies that energy flows out of the horizon, adding to the orbital energy and counteracting the loss to infinity. This ``energy return'' slows down the classical inspiral.}

{The black mirror absorbs the energy that the classical black hole would return. Therefore, the black mirror loses orbital energy at a higher rate than the classical black hole. Consequently, the observable signature of a high-spin black mirror is an inspiral that proceeds faster than the classical prediction.}

\subsection{Implications for the Cosmic Growth of Black Holes}
\label{sec.CosmicGrowth}

{The adoption of the thermodynamically consistent reflectivity $R = e^{-|\tilde{\omega}|/T_H}$ has important implications for the cosmic growth of black holes. In our stabilized model, superradiance is suppressed ($R \to 0$ for high spins). A rapidly spinning black mirror absorbs accretion flow angular momentum efficiently, just like a standard absorber, without generating a repulsive outward flux. Therefore, there is no barrier to spin-up. Black mirrors can grow to become the rapidly spinning supermassive objects observed in the universe ($a_* \to 1$).}

{The distinguishing feature of these high-spin objects is not their non-existence, but their ``darkness'' to gravitational waves: they do not amplify superradiant modes.}

\section{Conclusions}

The fundamental distinction between a classical black hole and a black mirror resides in the boundary conditions imposed at the event horizon. The classical paradigm demands perfect absorption, while the black mirror hypothesis implies a reflective boundary.

We demonstrate that the interaction of gravitational waves with this structure is governed by the requirement that the horizon is a quantum system in thermal equilibrium. {Physical causality and the requirement for a viscous dissipation term mandate that the reflectivity is given by $R = e^{-|\tilde{\omega}|/T_H}$ (Eq.~\ref{eq.Kerr_Reflectivity}).} We establish a unique universality: the solution is entirely agnostic to the specific, unknown quantum gravity interactions occurring at the boundary. The macroscopic reflectivity is dictated exclusively by the Hawking temperature, rendering the underlying microscopic details irrelevant.

Gravitational waves provide the tool to dynamically probe these boundary conditions. EMRIs are uniquely suited for testing this hypothesis. The long inspiral acts as a secular probe.

Our analysis demonstrates that the astrophysical implications are strongly dependent on the spin $a_*$ of the central object.

In the low-spin regime ($a_* < a_c$), the reflectivity causes a reduction in absorption ($R > 0$, but small). This leads to a secular dephasing where the black mirror inspiral is slower than the classical prediction.

In the high-spin regime ($a_* > a_c$), {the stabilization of the black mirror physics leads to a distinct conclusion. The black mirror suppresses the superradiant amplification that characterizes classical Kerr black holes. By acting as a perfect absorber in the superradiant regime, the black mirror drains orbital energy faster than a classical black hole. This leads to an inspiral that is faster than the classical prediction.}

{This dichotomy—slower inspiral at low spin, faster inspiral at high spin—provides a unique, falsifiable fingerprint for the black mirror hypothesis.} Furthermore, {the model is consistent with the existence of high-spin astrophysical black holes, as it allows for cosmic growth.}

Detection strategies will rely on matched filtering employing templates that incorporate the predicted generalized reflectivity $R(\omega, m)$, enabling a rigorous test of the black mirror hypothesis against the classical paradigm. The black mirror theory makes a concrete prediction that moves this beyond a purely academic exercise. Upcoming gravitational wave data will soon test these horizon properties, forcing us to confront the possibility that astrophysical black holes are not\ldots holes at all.

\section*{Acknowledgments}

We thank to Kostas Tzanavaris for his comments on the manuscript, for the many discussions we had on black mirrors, and for urging us to prove his black mirrors work wrong. We are indebted to Niayesh Afshordi and Naritaka Oshita for discussions about the reflectivity.

\end{document}